\documentstyle[aps,preprint]{revtex}
\input epsf
\global\firstfigfalse  

\tighten

\begin{document}
\pagestyle{empty}

\preprint{
\begin{minipage}[t]{3in}
\end{minipage}
}

\title{Statefinder diagnostic for holographic dark energy model}

\author{Xin Zhang \\\bigskip}
\address{Institute of High Energy Physics, Chinese Academy of Sciences\\
P.O.Box 918(4), Beijing 100049, People's Republic of China\\
\smallskip{\tt zhangxin@mail.ihep.ac.cn}}

\maketitle

\begin{abstract}

In this paper we study the holographic dark energy model proposed
by Li from the statefinder viewpoint. We plot the evolutionary
trajectories of the model with $c=1$ in the statefinder
parameter-planes. The statefinder diagrams characterize the
properties of the holographic dark energy and show the
discrimination between this scenario and other dark energy models.
We also perform a statefinder diagnostic to the holographic dark
energy model in cases of different $c$ which given by three fits
to observational data. The result indicates that from the
statefinder viewpoint $c$ plays a significant role in this model
and should thus be determined seriously by future high precision
experiments.

\vskip .4cm


\vskip .2cm


\end{abstract}

\newpage
\pagestyle{plain} \narrowtext \baselineskip=18pt

\setcounter{footnote}{0}

Physicists and astronomers begin to consider the dark energy
cosmology seriously and to explore the nature of dark energy
actively since the expansion of our universe is proven to be
accelerating at present time by the Type Ia supernovae
observations \cite{sn}. The analysis of cosmological observations,
in particular of the WMAP (Wilkinson Microwave Anisotropy Probe)
experiment \cite{wmap}, indicates that dark energy occupies about
2/3 of the total energy of our universe, and dark matter about
1/3. The accelerated expansion of the present universe is
attributed to that dark energy is an exotic component with
negative pressure, and many models have been constructed for
interpreting or describing this component. It should be pointed
out that the hottest candidates for dark energy are the
cosmological constant (or vacuum energy) \cite{cc} and the
quintessence scalar field \cite{quin}. However, as is well known,
there are two difficulties arise from all of these scenarios,
namely the two dark energy (or cosmological constant) problems --
the fine-tuning problem and the cosmic coincidence problem. The
fine-tuning problem asks why the dark energy density today is so
small compared to typical particle scales. The dark energy density
is of order $10^{-47} {\rm GeV}^4$, which appears to require the
introduction of a new mass scale 14 or so orders of magnitude
smaller than the electroweak scale. The second difficulty, the
cosmic coincidence problem, states: Since the energy densities of
dark energy and dark matter scale so differently during the
expansion of the universe, why are they nearly equal today? To get
this coincidence, it appears that their ratio must be set to a
specific, infinitesimal value in the very early universe.

Recently, considerable interest has been stimulated in explaining
the observed dark energy by the holographic dark energy model. For
an effective field theory in a box of size $L$, with UV cut-off
$\Lambda_c$ the entropy $S$ scales extensively, $S\sim
L^3\Lambda_c^3$. However, the peculiar thermodynamics of black
hole \cite{bh} has led Bekenstein to postulate that the maximum
entropy in a box of volume $L^3$ behaves nonextensively, growing
only as the area of the box, i.e. there is a so-called Bekenstein
entropy bound, $S\leq S_{BH}\equiv\pi M_p^2L^2$. This nonextensive
scaling suggests that quantum field theory breaks down in large
volume. To reconcile this breakdown with the success of local
quantum field theory in describing observed particle
phenomenology, Cohen et al. \cite{cohen} proposed a more
restrictive bound -- the energy bound. They pointed out that in
quantum field theory a short distance (UV) cut-off is related to a
long distance (IR) cut-off due to the limit set by forming a black
hole. In the other words, if the quantum zero-point energy density
$\rho_\Lambda$ is relevant to a UV cut-off, the total energy of
the whole system with size $L$ should not exceed the mass of a
black hole of the same size, thus we have $L^3\rho_\Lambda\leq
LM_p^2$, this means that the maximum entropy is in order of
$S_{BH}^{3/4}$. When we take the whole universe into account, the
vacuum energy related to this holographic principle
\cite{holoprin} is viewed as dark energy, usually dubbed
holographic dark energy. The largest IR cut-off $L$ is chosen by
saturating the inequality so that we get the holographic dark
energy density
\begin{equation}
\rho_\Lambda=3c^2M_p^2L^{-2}~,
\end{equation} where $c$ is a numerical constant, and $M_p\equiv 1/\sqrt{8\pi G}$ is the reduced Planck
mass. If we take $L$ as the size of the current universe, for
instance the Hubble scale $H^{-1}$, then the dark energy density
will be close to the observed data. However, Hsu \cite{hsu}
pointed out that this yields a wrong equation of state for dark
energy. Li \cite{li} subsequently proposed that the IR cut-off $L$
should be taken as the size of the future event horizon
\begin{equation}
R_h(a)=a\int_t^\infty{dt'\over a(t')}=a\int_a^\infty{da'\over
H(a')a'^2}~,\label{eh}
\end{equation} then the problem can be solved nicely and the
holographic dark energy model can thus be constructed
successfully. The holographic dark energy scenario may provide
natural solutions to both dark energy problems at the same time as
indicated in Ref.\cite{li}. Some speculations on the deep reasons
of the holographic dark energy were considered by several authors
\cite{holo1}. Further studies on this model can also be found in
Refs.\cite{holo2,snfit1,snfit2,cmb1,cmb2,cmb3,Nojiri}.

However, on the other hand, since more and more dark energy models
have been constructed for interpreting or describing the cosmic
acceleration, the problem of discriminating between the various
contenders is now emergent. In order to be able to differentiate
between those competing cosmological scenarios involving dark
energy, a sensitive and robust diagnostic for dark energy models
is a must. For this purpose a diagnostic proposal that makes use
of parameter pair $\{r,s\}$, the so-called ``statefinder", was
introduced by Sahni et al. \cite{sahni}. The statefinder probes
the expansion dynamics of the universe through higher derivatives
of the expansion factor $\stackrel{...}{a}$ and is a natural
companion to the deceleration parameter $q$ which depends upon
$\ddot a$. The statefinder pair $\{r,s\}$ is defined as follows
\begin{equation}
r\equiv
\frac{\stackrel{...}{a}}{aH^3},~~~~s\equiv\frac{r-1}{3(q-1/2)}~.\label{rs}
\end{equation} The statefinder
is a ``geometrical'' diagnostic in the sense that it depends upon
the expansion factor and hence upon the metric describing
space-time.

Trajectories in the $s-r$ plane corresponding to different
cosmological models exhibit qualitatively different behaviors. The
spatially flat LCDM (cosmological constant $\Lambda$ with cold
dark matter) scenario corresponds to a fixed point in the diagram
\begin{equation}
\{s,r\}\bigg\vert_{\rm LCDM} = \{ 0,1\} ~.\label{lcdm}
\end{equation}
Departure of a given dark energy model from this fixed point
provides a good way of establishing the ``distance'' of this model
from LCDM \cite{sahni,alam}. As demonstrated in Refs.
\cite{sahni,alam,gorini,zimdahl,zx} the statefinder can
successfully differentiate between a wide variety of dark energy
models including the cosmological constant, quintessence, the
Chaplygin gas, braneworld models and interacting dark energy
models. We can clearly identify the ``distance'' from a given dark
energy model to the LCDM scenatio by using the $r(s)$ evolution
diagram.

The current location of the parameters $s$ and $r$ in these
diagrams can be calculated in models, and on the other hand it can
also be extracted from data coming from SNAP (SuperNovae
Acceleration Probe) type experiments. Therefore, the statefinder
diagnostic combined with future SNAP observations my possibly be
used to discriminate between different dark energy models.

In this paper we apply the statefinder diagnostic to the
holographic dark energy model. The statefinder can also be used to
diagnose different cases of the model. Analysis of the
observational data provides constraints on the holographic dark
energy model. A direct comparison of the present available Type Ia
supernovae data with the holographic dark energy model was
undertaken recently \cite{snfit1}. Constraints on the parameters
of the model arising from the observations of cosmic microwave
background (CMB) was studied \cite{cmb1}. The possible connection
between the holographic dark energy and the low-$l$ CMB multipoles
was analyzed \cite{cmb2,cmb3}. However, it is remarkable that the
values of the free parameter $c$ determined by different authors
are so different. We use the statefinder to diagnose these
different cases, and the result shows that the parameter $c$ plays
an important role in the model and thus should be considered
seriously and should be determined accurately by future high
precision experiments.

In what follows we will calculate the statefinder parameters for
the holographic dark energy model and plot the evolutionary
trajectories of the model in the statefinder parameter-planes. We
now consider a spatially flat FRW (Friedmann-Robertson-Walker)
universe with matter component $\rho_m$ (including both baryon
matter and cold dark matter) and holographic dark energy component
$\rho_\Lambda$, the Friedmann equation reads
\begin{equation}
1=\Omega_m+\Omega_\Lambda~,
\end{equation} where $\Omega_m=\Omega_{m}^0H_0^2H^{-2}a^{-3}$ and
$\Omega_\Lambda=c^2H^{-2}R_h^{-2}$ are relative energy densities
of matter and dark energy, respectively, expressed as fractions of
the critical density $\rho_c=3M_p^2H^2$. By using
$R_h=c/H\sqrt{\Omega_\Lambda}$ and the definition of the event
horizon (\ref{eh}), we get
\begin{equation}
\int_x^\infty{dx'\over H(a')a'}={c\over
H(a)a\sqrt{\Omega_\Lambda}}~,\label{rh}
\end{equation} where $x=\ln a$. We notice that the Friedmann
equation implies
\begin{equation}
{1\over H(a)a}=\sqrt{a(1-\Omega_\Lambda)}{1\over
H_0\sqrt{\Omega_m^0}}~.\label{fri}
\end{equation} Combining (\ref{rh}) and (\ref{fri}), one obtains
the relation
\begin{equation}
\int_x^\infty e^{x'/2}\sqrt{1-\Omega_\Lambda}dx'=c
e^{x/2}\sqrt{{1\over\Omega_\Lambda}-1}~.
\end{equation} Then taking derivative with respect to $x$ in both
sides of the above relation, we get
\begin{equation}
\Omega_\Lambda '=\Omega_\Lambda(1-\Omega_\Lambda)(1+{2\over
c}\sqrt{\Omega_\Lambda})~, \label{deq}\end{equation} where the
prime denotes the derivative with respect to $x$. This
differential equation describes the behavior of the holographic
dark energy completely, and it can be solved analytically
\cite{li,snfit1}.

From the energy conservation equation of the dark energy, the
equation of state of the dark energy can be expressed as
\begin{equation}
w=-1-{1\over 3}{d\ln\rho_\Lambda\over d\ln a}~.
\end{equation} Then making use of the formula
$\rho_\Lambda={\Omega_\Lambda\over
1-\Omega_\Lambda}\rho_m^0a^{-3}$ and the differential equation of
$\Omega_\Lambda$ (\ref{deq}), the equation of state for the
holographic dark energy can be given
\begin{equation}
w=-{1\over 3}(1+{2\over c}\sqrt{\Omega_\Lambda})~.\label{w}
\end{equation}
Also, we can give the change rate of $w$ with respect to $x=\ln
a$,
\begin{equation}
w'=-{1\over 3c}\sqrt{\Omega_\Lambda}(1-\Omega_\Lambda)(1+{2\over
c}\sqrt{\Omega_\Lambda})~.\label{w'}
\end{equation} It can be seen clearly that the equation of state
of the holographic dark energy satisfies $-(1+2/c)/3\leq w\leq
-1/3$ due to $0\leq\Omega_\Lambda\leq 1$. If we take $c=1$, the
behavior of the holographic dark energy will be more and more like
a cosmological constant with the expansion of the universe, and
the ultimate fate of the universe will be entering the de Sitter
phase in the far future. As is shown in Ref.\cite{li}, if one puts
the parameter $\Omega_\Lambda^0=0.73$ into (\ref{w}), then a
definite prediction of this model, $w^0=-0.903$, will be given.

The statefinder parameters $r$ and $s$ can also be expressed as
\begin{equation}
r=1+{9\over 2}w(1+w)\Omega_\Lambda-{3\over 2}w'\Omega_\Lambda~,
\end{equation}
\begin{equation}
s=1+w-{1\over 3}{w'\over w}~.
\end{equation} Thus, using (\ref{w}) and (\ref{w'}) we can give
these parameters in terms of $\Omega_\Lambda$,
\begin{equation}
r=1-\Omega_\Lambda-{1\over
2c}\Omega_\Lambda^{3/2}(1+\Omega_\Lambda)+{1\over
c^2}\Omega_\Lambda^2(3-\Omega_\Lambda)~,
\end{equation}
\begin{equation}
s={2\over 3}-{1\over 3c}\sqrt{\Omega_\Lambda}(3-\Omega_\Lambda)~.
\end{equation} Another important parameter in the statefinder
diagnostic is the deceleration parameter $q=-\ddot{a}/aH^2$, we
also express it in terms of $\Omega_\Lambda$,
\begin{equation}
q={1\over 2}-{1\over 2}\Omega_\Lambda-{1\over
c}\Omega_\Lambda^{3/2}~.
\end{equation} Though the relations between these statefinder parameters,
namely the functions $r(s)$ and $r(q)$, can be derived
analytically in principle, we do not give the expressions here due
to the complexity of the formulae. Making the parameter
$\Omega_\Lambda$ vary from 0 to 1, one can easily get the
evolutionary trajectories in the statefinder parameter-planes of
this model. As an example, we plot the statefinder diagrams in the
$s-r$ plane and $q-r$ plane corresponding to the case $c=1$ in
Fig.1 and Fig.2.

\vskip.8cm
\begin{figure}
\begin{center}
\leavevmode \epsfbox{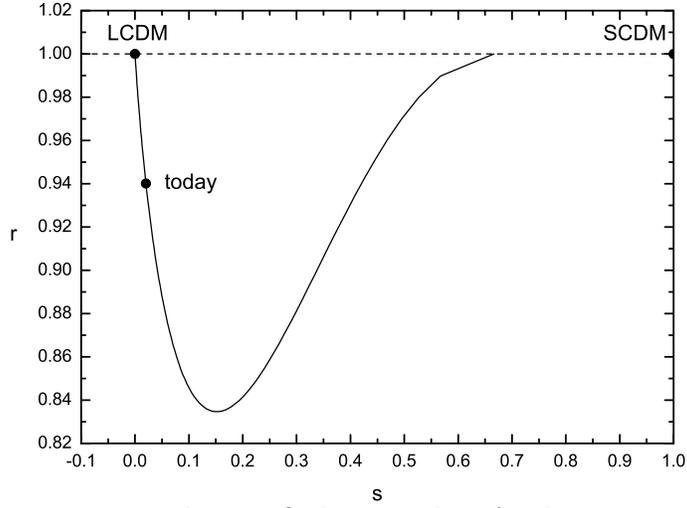} \caption[]{Evolution trajectory in
the statefinder $s-r$ plane for the case $c=1$. LCDM corresponds
to a fixed point $(0,1)$, and SCDM corresponds to $(1,1)$. For the
holographic dark energy model, $s$ monotonically decreases from
2/3 to 0, whereas $r$ first decreases from 1 to a minimum value,
then rises to 1. The coordinate of today's point is $(0.02,
0.94)$, thus the 'distance' from this model to the LCDM can be
easily identified in this diagram.}
\end{center}
\end{figure}

\vskip.8cm
\begin{figure}
\begin{center}
\leavevmode \epsfbox{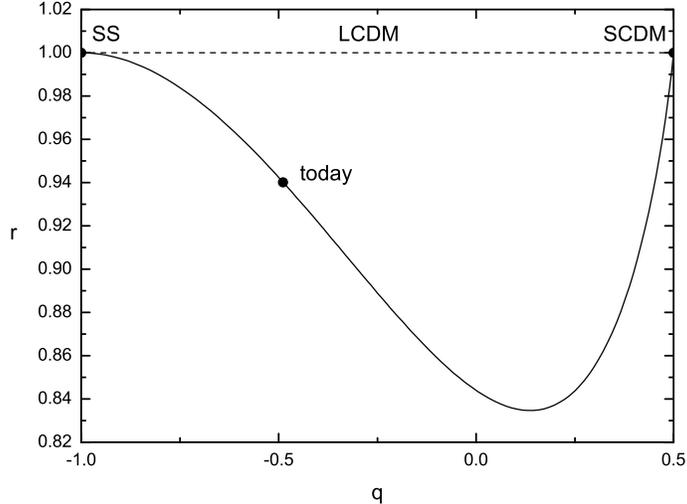} \caption[]{Evolution trajectory in
the statefinder $q-r$ plane for the case $c=1$. The solid line
represents the holographic dark energy model, and the dashed line
the LCDM as comparison. Both models diverge at the same point in
the past $(0.5,1)$ which corresponds to a matter dominated
universe (SCDM), and converge to the same point in the future
$(-1,1)$ which corresponds to the steady state model (SS) - the de
Sitter expansion. The point of today in this plane locates at
$(-0.49, 0.94)$. }
\end{center}
\end{figure}

The statefinder diagnostic can discriminate between various dark
energy models effectively. Different cosmological models involving
dark energy exhibit qualitatively different evolution trajectories
in the $s-r$ plane. For example, the LCDM scenario corresponds to
the fixed point $s=0,~r=1$ as shown in (\ref{lcdm}), and the SCDM
(standard cold dark matter) scenario corresponds to the point
$s=1,~r=1$. For the so-called ``quiessence" models ($w$ is a
constant), the trajectories are some vertical segments, i.e. $r$
decreases monotonically from 1 to $1+{9\over 2}w(1+w)$ while $s$
remains constant at $1+w$ \cite{sahni}. The quintessence (inverse
power law) tracker models and the Chaplygin gas models have
typical trajectories similar to arcs of a parabola (upward and
downward) lying in the regions $s>0,~r<1$ and $s<0,~r>1$,
respectively \cite{sahni,alam,gorini}. The coupled quintessence
models exhibit more complicated trajectories as shown in Ref.
\cite{zx}. For the holographic dark energy scenario, as shown in
Fig.1 ($c=1$ case), commences its evolution from $s=2/3,~r=1$,
through an arc segment, and ends it at the LCDM fixed point
($s=0,~r=1$) in the future. We see clearly from this diagram that
$s$ decreases monotonically to zero, while $r$ first decreases to
a minimum value then increases to unity. According to the
holographic dark energy model, the current universe corresponds to
a point $s=0.02,~r=0.94$ in the plane. Therefore, the ``distance''
from this model to the LCDM scenario can be identified explicitly.
The distinctive trajectories which various dark energy scenarios
follow in the $s-r$ plane demonstrate quite strikingly the
contrasting behavior of dark energy models.

As a complementarity, Fig.2 shows another statefinder diagram ---
the $r(q)$ evolutionary trajectory. From this figure, we clearly
see that both LCDM scenario and holographic dark energy model
commence evolving from the same point in the past $(q=0.5, r=1)$
which corresponds to a matter dominated SCDM universe, and end
their evolution at the same common point in the future $(q=-1,
r=1)$ which corresponds to a steady state cosmology (SS) --- the
de Sitter expansion. The trajectories of LCDM model and
holographic dark energy model follow are horizontal segment and
arc segment, respectively, in this diagram. In the holographic
dark energy model, the point of today in this plane locates at
$q=-0.49, r=0.94$.

Theoretically, it has been argued in Ref.\cite{li} that, for the
numerical parameter, $c=1$ is a good choice. However, one prefers
to consider $c$ as a free parameter of the model when performing a
fit to observational data. Actually, if $c<1$, the holographic
dark energy will behave like a Quintom proposed recently in
Ref.\cite{quintom}, the amazing feature of which is that the
equation of state of dark energy component $w$ crosses $-1$, i.e.
it is larger than $-1$ in the past while less than $-1$ near
today. The recent fits to current SN Ia data with parametrization
of the equation of state of dark energy find the Quintom type dark
energy is mildly favored \cite{running}. Usually the Quintom dark
energy model is realized in terms of double fields -- a
quintessence and a phantom, which dominate at earlier time and
today respectively \cite{relev,quintom1}. However, the holographic
dark energy in the case $c<1$ provides a more natural realization
of the Quintom picture. While, if $c>1$, the equation of state of
dark energy will be always larger than $-1$ such that the universe
avoids entering the de Sitter phase and the Big Rip phase. Hence,
we see explicitly, the determining of the value of $c$ is a key
point to the feature of the holographic dark energy and the
ultimate fate of the universe as well. However, in the recent fit
studies, different groups gave different values to $c$. A direct
fit of the present available SNe Ia data with this holographic
model indicates that the best fit result is $c=0.21$
\cite{snfit1}. Recently, by calculating the average equation of
state of the dark energy and the angular scale of the acoustic
oscillation from the BOOMERANG and WMAP data on the CMB to
constrain the holographic dark energy model, the authors show that
the reasonable result is $c\sim 0.7$ \cite{cmb1}. In addition, the
holographic dark energy model has also been used to investigate
the suppression of the power at low multipoles in the CMB spectrum
\cite{cmb2}. In the study of the constraints on the dark energy
from the holographic connection to the small $l$ CMB suppression,
an opposite result is derived, i.e. it implies the best fit result
is $c=2.1$ \cite{cmb3}. We see explicitly that these results are
very incompatible. In order to demonstrate the tremendous
difference between these cases in the time evolution of the
universe, we perform a statefinder diagnostic to these cases.

\vskip.8cm
\begin{figure}
\begin{center}
\leavevmode \epsfbox{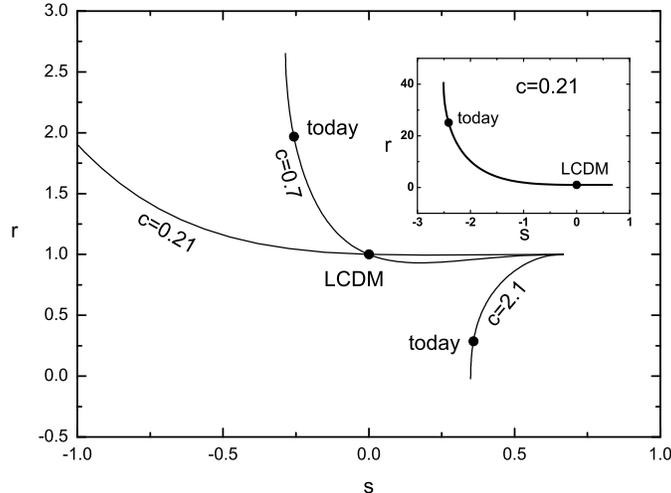} \caption[]{The statefinder diagrams
$r(s)$ for holographic dark energy model in the cases of
$c=0.21,~0.7$ and 2.1, respectively. The inset shows the complete
curve for the case $c=0.21$. The coordinates of today are $(-2.41,
25.13)$, $(-0.26, 1.97)$ and $(0.36, 0.29)$, corresponding to the
cases $c=0.21,~0.7$ and 2.1, respectively. }
\end{center}
\end{figure}

We plot in Fig.3 the statefinder diagrams $r(s)$ for the
holographic dark energy model in three cases of which the
parameter $c$ is taken to be $0.21,~0.7$ and 2.1, respectively.
This figure strikingly demonstrates that the evolution
trajectories of these cases in the $s-r$ plane are tremendously
different. We see that the evolutionary trends of cases $c<1$ and
$c>1$ are upward and downward, respectively. In particular, we
notice that for the $c=0.21$ case, $r$ can arrive at a very large
value, about 40, as shown in the inset in Fig.3. For the
situations of $c<1$, the trajectories pass through the LCDM fixed
point; while for the $c>1$ cases, the tracks never reach the LCDM
fixed point. The coordinates of today's statefinders are $(-2.41,
25.13)$, $(-0.26, 1.97)$ and $(0.36, 0.29)$, corresponding to the
cases $c=0.21,~0.7$ and 2.1, respectively. From this figure it can
be seen that the model universes at today corresponding to
$c=0.21$ and $c=0.7$ have evolved through the LCDM fixed point. It
is clear that the present values of the statefinders, $r$ and $s$,
can be used to distinguish the holographic dark energy models with
different values of $c$. Obviously, the distances between these
cases can also be easily measured. It is remarkable that, from the
statefinder viewpoint, the parameter $c$ plays an important role
in this model, and it determines the evolution behavior as well as
the ultimate fate of the model universe. We hope that the future
high precision experiments (e.g. SNAP) may provide sufficiently
large amount of precise data to be capable of determining the
value of $c$.

In summary, we study in this paper the holographic dark energy
model from the statefinder viewpoint. We plot the evolutionary
trajectories of this model in the statefinder parameter-planes.
The statefinder diagrams characterize the properties of the
holographic dark energy and show the discrimination between this
model and other dark energy models. The statefinder diagnostic can
also be performed to the holographic dark energy model in cases of
different $c$, in this paper we chose three fit results as
example, which indicates that the value of $c$ determines the
evolution behavior and the fate of the universe. We hope that the
future large amount of data on high-$z$ Type Ia supernovae may be
capable of determining the statefinder parameters and consequently
shed light on the nature of dark energy.

\begin{acknowledgments}
This work was supported in part by the Natural Science Foundation
of China (Grant No. 10375072).
\end{acknowledgments}



\end{document}